\documentclass[journal]{IEEEtran}
\normalsize

\usepackage{cite}
\ifCLASSINFOpdf
   \usepackage[pdftex]{graphicx}
   \DeclareGraphicsExtensions{.pdf,.jpeg,.png}
\else
   \usepackage[dvips]{graphicx}
   \DeclareGraphicsExtensions{.eps}
\fi
\usepackage[cmex10]{amsmath}
\usepackage{amssymb}
\usepackage{color}

\usepackage[normalem]{ulem}

\DeclareMathOperator*{\argmax}{\arg\!\max}

\begin{document}
\title{Low Complex, Narrowband-Interference Robust Synchronization for NC-OFDM Cognitive Radio}

\author{Pawel~Kryszkiewicz,~\IEEEmembership{Student Member,~IEEE,}
        and~Hanna~Bogucka,~\IEEEmembership{Senior Member,~IEEE}% <-this % stops a space
\thanks{The authors are with the Chair of Wireless Communications, Poznan University of Technology, Poznan, Poland e-mail: pawel.kryszkiewicz@put.poznan.pl.}% <-this % stops a space
% <-this % stops a space
\thanks{Copyright (c) 2016 IEEE. Personal use is permitted. For any other purposes, permission must be obtained from the IEEE by emailing pubs-permissions@ieee.org.
This is the author's version of an article that has been published in this journal. Changes were made to this version by the publisher prior to publication.
The final version of record is available at http://dx.doi.org/10.1109/TCOMM.2016.2596780}% 
\thanks
{The work presented in this paper has been implemented within the PRELUDIUM project funded by the National Science Centre, Poland, based on decision no. DEC-2012/05/N/ST7/00164 and by the Polish Ministry of Science and Higher Education within the status activity task \emph{Cognitive radio systems} in 2016.}
%\thanks{TBD}
% <-this % stops a space
%\thanks{Manuscript received April 19, 2005; revised December 27, 2012.}
}


\markboth{IEEE Transactions on Communications}%
{Submitted paper}



\maketitle

\begin{abstract}
This paper presents a new, low-complexity algorithm for time and frequency synchronization in a Non-Contiguous-Orthogonal Frequency Division Multiplexing (NC-OFDM) radio communication system in the presence of narrowband interference (NBI). This interference scenario is plausible for Cognitive Radio (CR) systems. The computational complexity of the proposed synchronization method is similar to the known Schmidl\&Cox algorithm. The algorithm is partially blind, i.e., it does not require information on the NC-OFDM subcarriers used, nor on the center-frequency of the NBI. Moreover, it does not require filtering to remove the NBI. It performs best for the NBI of constant carrier frequency, but it also provides good results for the frequency-modulated (FM) interfering signal.      

%{\PK nazwa: Narrowband interferer robust synchronization- (NIRS)?? LOW COmplexity? Energy Efficient?}
\end{abstract}


% \fi
\IEEEpeerreviewmaketitle

\section{Introduction}
\IEEEPARstart{N}{on}-Contiguous-Orthogonal Frequency Division Multiplexing (NC-OFDM) is a modulation and multiplexing technique considered for Cognitive Radio (CR) systems. It enables the aggregation of fragmented spectrum \cite{Mahmoud09_OFDM_for_CR}, and protect of incumbent systems referred to as Primary Users (PUs). In an NC-OFDM transmitter, data symbols modulate only the subcarriers (SCs) that do not coincide with the PU's band. The other ones are modulated by zeros in order to limit the interference power observed at the PU receiver. The often considered CR use-case is in an UHF band, used by digital TV and narrowband wireless microphones \cite{Mwangoka_2011_COGEU}. These systems' receivers have to be protected from the interference introduced by CR transmission. A number of techniques have been proposed for NC-OFDM-based CR in order to limit the generated interference power, e.g., adaptive SC power allocation \cite{Bansal11_przydzial_mocy}, reservation of guard bands, or some advance spectrum-shaping methods \cite{Kryszkiewicz_OCCS_2013}. However, no polices are in place to protect CR transmission. Therefore, CRs have to be designed to deal with interference originating from the licensed systems. An important design issue, in this case, is an interference-robust synchronization algorithm.  

A low-complexity synchronization method for OFDM is the preamble-based Schmidl\&Cox (S\&C) algorithm \cite{Schmidl_Cox_1997}. It can be adopted to NC-OFDM by using only the available subcarriers (outside the PUs band) for the preamble. However, at the NC-OFDM receiver, high-power PUs-generated interference is expected. Importantly, even strong interference may not deteriorate NC-OFDM data symbols detection, because this detection is implemented in the frequency domain (at the output of DFT block). However, before symbol detection, the first synchronization phase must always be implemented in the time domain, in which unknown interference cannot be easily removed. There have been some studies on the S\&C algorithm performance in the presence of various interference models. Wideband interference degrades its performance similarly to white noise \cite{Zivkovic_2011_synchr_SC_wideband}, however, narrowband interference (NBI) is more harmful. As shown in \cite{Coulson_2004_synchr_narrow_sinusoid,Marey_2007_SC_narrowband}, autocorrelation-based S\&C synchronization mistakes NBI for a preamble, causing the so-called \emph{false synchronization}. Although filtering can be applied to remove NBI, it involves computationally-complex detection of the NBI frequency, the filter design and filtering itself \cite{Coulson_2004_synchr_narrow_sinusoid}. Another approach is to base synchronization on cross-correlation, as in Ziabari's method utilizing its own cross-correlation-like scheme \cite{Ziabari_synchr_NC_OFDM} at the cost of high complexity. In \cite{Sanguinetti_2010_synchr}, a synchronization algorithm for OFDM robust against NBI is presented. Unfortunately, it has a very high computational complexity and requires a special preamble shape consisting of 4 identical sequences. The third sequence has an inverted sign, which increases interference leakage into the PU band. As such, it is typically not suitable for NC-OFDM-based CR.
  
A new synchronization scheme proposed in this paper aims at removing NBI impact on synchronization metrics at a relatively small computational cost over the S\&C algorithm. This new Narrowband-Interference Robust Synchronization (NIRS) algorithm is partially blind, i.e., it does not require \emph{a priori} knowledge on the NC-OFDM occupied SCs, nor on the NBI center frequency. It assumes that the absolute value of the Carrier Frequency Offset (CFO) does not exceed the NC-OFDM SC spacing.
Although this assumption limits the CFO estimation range of NIRS, it seems to be practical, considering contemporary hardware and practical CR use-cases. Let us consider an LTE-like NC-OFDM-based CR system operating at the carrier frequency of 700~MHz. The maximum assumed User Equipment frequency error is 70~Hz, because the required frequency stability of the user terminal is 0.1~ppm \cite{3gpp.36.101}. For the worst case of a Home Base Station this error equals 175 Hz due to a frequency stability requirement of 0.25~ppm \cite{3gpp.36.104}. Considering the maximum frequency shift in a fast-train scenario (moving with  350~km/h) caused by the Doppler effect of 227~Hz, the maximum CFO is 472~Hz. Because SC spacing in an LTE system is 15~kHz, the fractional CFO estimation range of (--15~kHz; 15~kHz) in the proposed method is sufficient.
	
In Section II, we present the S\&C algorithm applied to NC-OFDM. In Section III-IV, we present our new NIRS algorithm and evaluate its performance. Section V contains conclusions.

\section{Schmidl\&Cox Method for NC-OFDM}
%\label{sec:system_model}
The considered NC-OFDM system uses an $N$-point (Inverse) Discrete Fourier Transform (IDFT/DFT) as the modulator/demodulator. Each NC-OFDM symbol is preceded by $N_{\mathrm{CP}}$ samples of the Cyclic Prefix (CP). The $n$-th sample of the $p$-th NC-OFDM symbol equals:
\begin{equation}
x^{(p)}_{n}\!\!=\!\!\left\{\!\!\!\!
\begin{array}{ll}
\frac{1}{\sqrt{N}}
\sum_{k=-N/2}^{N/2-1}\!d_{k}^{(p)}e^{j2\pi \frac{nk}{N}} \!\!& \!\!\mbox{for $\!-N_{\mathrm{CP}}\!\leq\! n\!\!\leq\!\! N\!-\!\!1$}\\
0 & \!\!\mbox{otherwise,}
\end{array}
\right.
\label{eq_IFFT}
\end{equation}
where $d_{k}^{(p)}$ is a complex symbol transmitted on the $k$-th SC in the $p$-th symbol. Importantly, only $\alpha$ out of $N$ SCs in the NC-OFDM transmitter are modulated by non-zero QAM/PSK symbols. The indices of occupied SCs constitute a vector $\mathbf{I}=\{I_{j}\}$ for $j=1,...,\alpha$ and $I_{j}\in \{-N/2,..., N/2-1\}$. Released SCs whose indices are not in $\mathbf{I}$ are zeroed in order to protect licensed transmission, i.e., $d_{k}^{(p)}=0$ for $k\notin\mathbf{I}$.   
The signal is transmitted in frames, each consisting of $P$ NC-OFDM symbols.The discrete-time NC-OFDM frame is defined as the concatenation of NC-OFDM symbols:
\begin{equation}
\tilde{x}(n)=\sum_{p=0}^{P-1} x_{n-p(N+N_{\mathrm{CP}})}^{(p)}.
\end{equation}

The signal at the input of the NC-OFDM receiver, distorted by multipath fading and CFO is:  
\begin{equation}
r(n)=y(n)e^{\jmath 2\pi \frac{\nu n}{N}}+\sqrt{\sigma^{2}_{\mathrm{i}}}e^{\jmath 2\pi \frac{f_{\mathrm{c}} n}{N}+\jmath \varphi_{n}}+w(n), 
\label{eq:signal_model}
\end{equation}
where 
\begin{equation}
y(n)=\sum_{l=0}^{L-1}\tilde{x}(n-l)h(l)
\label{eq:channel_influence}
\end{equation}
is the received NC-OFDM signal distorted by the multipath channel, $h(l)$ is the $l$th path channel coefficient, $L$ is the number of channel paths components, $\nu$ is the CFO normalized to SCs spacing, and $w(n)$ is a complex white Gaussian noise with zero mean and variance $\sigma^{2}_{\mathrm{w}}$. The second component in (\ref{eq:signal_model}) is the NBI characterized by power $\sigma^{2}_{\mathrm{i}}$, center frequency $f_{\mathrm{c}}$ normalized to SCs spacing and a slowly-varying time-continuous phase whose value at the $n$th sampling moment is $\varphi_{n}$. Importantly, this NBI model is valid for many commonly used narrowband systems, e.g., wireless microphones (using frequency modulation) or GSM (using Gaussian Minimum Shift Keying) that have a constant amplitude and continuously varying phase, which can be described using the polynomial:
\begin{equation}
\varphi_{n}=\sum_{i=0}^{\infty}a_{i}n^{i},
\label{eq:phase_of_NBI}
\end{equation}
where  $a_{i}$ are the polynomial coefficients. Assuming that the NBI phase variations are slow in time (because the NBI bandwidth is narrow) and that the sampling frequency is many times higher than the NBI bandwidth, $N$ consecutive samples can be approximated by the linear function, i.e., $a_{i}$ for $i\in\{0,1\}$ being possibly non-zero for a given $n$. Thus, equation (\ref{eq:signal_model}) can be rewritten as
\begin{align}
r(n)&=\!y(n)e^{\jmath 2\pi \frac{\nu n}{N}}\!+\!\sqrt{\sigma^{2}_{\mathrm{i}}}e^{\jmath 2\pi \frac{f_{\mathrm{c}} n}{N}+\jmath (a_{0}+a_{1}n)}+w(n)
\nonumber\\&
\!=\!y(n)e^{\jmath 2\pi \frac{\nu n}{N}}\!+\!\sqrt{\sigma^{2}_{\mathrm{i}}}e^{\jmath 2\frac{\pi}{N}\left(f_{\mathrm{c}}+\frac{a_{1}N}{2\pi} \right)n+\jmath a_{0}}
+w(n)
\nonumber\\&
=y(n)e^{\jmath 2\pi \frac{\nu n}{N}}\!+\!\sqrt{\sigma^{2}_{\mathrm{i}}}e^{\jmath 2\pi \frac{f n}{N}+\jmath\varphi_{0}} +w(n),
\label{eq:signal_model2}
\end{align}
where 
\begin{equation}
f=f_{\mathrm{c}}+\frac{a_{1}N}{2\pi}
\end{equation}
is the NBI instantaneous frequency normalized to SCs spacing and $\varphi=a_{0}$ is the NBI signal phase. Importantly, this model is valid for a limited number of samples around a given $n$ value. For the sake of simplicity, this case is assumed in the following formulas, because the maximum of $N$ consecutive samples is always taken into account, i.e., this model is valid locally around a given $n$, although $f$ and $\varphi$ can be different for much different time instances $n$. Obviously, as the bandwidth of interference increases, this model becomes less accurate. Note that the same NBI model has been used in \cite{Coulson_2004_synchr_narrow_sinusoid}.   

The S\&C method uses the first NC-OFDM symbol in each frame as a preamble consisting of two identical time-domain sequences of $N/2$ samples. Only evenly indexed SCs from $\mathbf{I}$ are modulated in the preamble. Moreover, 
preamble samples are multiplied by $\sqrt{2}$ to maintain the same power as for data-bearing symbols. Importantly, the multipath effect does not destroy this repeatability, thanks to the CP. The autocorrelation operation used in the S\&C algorithm is defined as:
\begin{equation}
G(n)=\sum_{m=0}^{N/2-1}r^{*}(n+m)r(n+N/2+m)\; ,
\label{eq:simple_autocorrelation}
\end{equation}     
where $G(n)$ is the autocorrelation result, and $(~)^{*}$ denotes a complex conjugate. In the case of a no-noise and no-interference scenario, the maximum of $|G(n)|$ should be found for $n\in\{-N_{CP}+L-1,...,0\}$  (assuming that $L\leq N_{\mathrm{CP}}$). The received signal energy $M(n)$ 
%of $G(n)$ 
over $N/2$ samples equals:
\begin{equation}
M(n)=\sum_{m=0}^{N/2-1}|r(n+N/2+m)|^{2}.
\label{eq:simple_autocorrelation_denominator}
\end{equation}   
In the S\&C algorithm, the beginning of a frame is decided for $n$ maximizing the timing metric defined as $|G(n)/M(n)|^{2}$:

\begin{equation}
\widehat{n}= \argmax_{n} \left|\frac{G(n)}
{M(n)}\right|^{2}\; .
\label{eq:decision_metric_S_C}
\end{equation}  
As shown in \cite{Schmidl_Cox_1997}, $G(n)$ and $M(n)$ can be calculated iteratively (based on $G(n-1)$ and $M(n-1)$) with low complexity. Another option proposed in \cite{Schmidl_Cox_1997} is to find $\widehat{n}$ as a time-point lying in the middle between two points (closest, on each side) achieving 90\% of the maximum found using (\ref{eq:decision_metric_S_C}). 

The CFO estimate can be obtained as
\begin{equation}
\widehat{\nu}=\arg \left\{ \frac{G(\widehat{n})}{\pi} \right\}\; ,
\label{eq:decision_metric_S_C_frac_frequency}
\end{equation}      
where $\arg\{\cdot\}$ is the argument of a complex number. Note that this estimate is limited to the range of $(-1;1)$ of SC spacing, which is a typical range of the CFO considered (as argued in the Introduction for an LTE-like CR system, which is also assumed in the simulations discussed in Section \ref{sec:Simulations}).

\subsection{Autocorrelation effect in the case of NBI domination over noise}
\label{sec:characteristic_S_C_NBI}
Let us first consider an NC-OFDM system in which NBI plays the dominant role, and the noise can be neglected. This is a plausible scenario for an NC-OFDM-based CR. Assuming that an NC-OFDM system operates in a relatively high SNR range, the white noise component can be temporarily neglected in (\ref{eq:signal_model2}),\footnote{A statistical description of the S\&C algorithm in an OFDM system in the presence of NBI and noise can be found in \cite{Marey_2007_SC_narrowband}.} and by substituting (\ref{eq:signal_model2}) to (\ref{eq:simple_autocorrelation}), we obtain: 
\begin{equation}
G(n)\approx G_{\mathrm{y}}(n)+G_{\mathrm{i}}(n)+G_{\mathrm{cross}}(n),
\label{eq:autocorrelation_components}
\end{equation}
where
\begin{equation}
G_{\mathrm{y}}(n)=e^{\jmath \pi \nu}\sum_{m=0}^{N/2-1}y^{*}(n+m)y\left(n+m+\frac{N}{2}\right),
\label{eq:autocorrelation_components_G_x}
\end{equation}
\begin{equation}
G_{\mathrm{i}}(n)=\sum_{m=0}^{N/2-1}\sigma^{2}_{\mathrm{i}}e^{\jmath \pi f}=\frac{N}{2}\sigma^{2}_{\mathrm{i}}e^{\jmath \pi f},
\end{equation}
\begin{align}
\label{Gcross}
\!\!G_{\mathrm{cross}}(n)\!&\!=\!e^{\jmath \pi f+\jmath\varphi}\sqrt{\sigma^{2}_{\mathrm{i}}}\!\sum_{m=0}^{N/2-1}\!\!y^{*}(n\!+\!m)
e^{\jmath 2 \pi(f-\nu)\frac{n+m}{N}}
\!\!+\!
%\nonumber
\\&
\!e^{\jmath \pi \nu-\jmath\varphi}\!\sqrt{\sigma^{2}_{\mathrm{i}}}\!\sum_{m=0}^{N/2-1}\!\!\!y\!\left(\!n\!+\!m\!+\!\frac{N}{2}\right)
e^{-\jmath 2 \pi(f-\nu)\frac{n+m}{N}}
\nonumber\\&
=
e^{\jmath \pi f+\jmath\varphi}\sqrt{\sigma^{2}_{\mathrm{i}}}b^{*}(n)+\!e^{\jmath \pi \nu-\jmath\varphi}\sqrt{\sigma^{2}_{\mathrm{i}}}b\left(n+\frac{N}{2} \right),
\nonumber
\end{align}
%\begin{align}
%G_{\mathrm{iy}}(n)\!&=\!e^{\jmath \pi \nu-\jmath\varphi}\sqrt{\sigma^{2}_{\mathrm{i}}}\sum_{m=0}^{N/2-1}\!y\!\left(\!n\!+\!m\!+\!\frac{N}{2}\right)
%e^{-\jmath 2 \pi(f-\nu)\frac{n+m}{N}}
%\nonumber \\&
%=\!e^{\jmath \pi \nu-\jmath\varphi}\sqrt{\sigma^{2}_{\mathrm{i}}}b\left(n+\frac{N}{2} \right),
%\label{Giy}
%\end{align}
\begin{align}
b(n)=e^{-\jmath 2 \pi(f-\nu)\frac{n}{N}}\sum_{m=0}^{N/2-1}y\left(m\right)
e^{-\jmath 2 \pi(f-\nu)\frac{n+m}{N}}.
\label{eq:DFT_z_cross}
\end{align}

In the above, $G_{\mathrm{y}}(n)$ denotes the NC-OFDM signal autocorrelation, $G_{\mathrm{i}}(n)$ denotes the NBI autocorrelation and $G_{\mathrm{cross}}(n)$ results from the cross-correlation between the NBI and the NC-OFDM received signals. Importantly, the term $G_{\mathrm{cross}}(n)$ can be expressed using phase-rotated results of the Discrete Fourier Transform (DFT) of the received NC-OFDM signal $b(n)$ defined by (\ref{eq:DFT_z_cross}). Because NC-OFDM SCs of frequencies adjacent to NBI frequency are inactive, i.e., modulated by zeros ($d_{k}^{(p)}=0$ for $k\approx f$), and because CFO $\nu$ is assumed to be relatively small ($|\nu|<1$), the values of $b(n)$ and $b\left(n+\frac{N}{2}\right)$ (and thus also $G_{\mathrm{cross}}(n)$) in the NC-OFDM should be small in comparison with the case for a standard OFDM system. This is confirmed analytically and on the basis of the examples in Annex \ref{sec:appendix}.

This means that for an NC-OFDM signal separated in frequency from the NBI,\footnote{This separation may also involve the application of guard bands (guard SCs).} the autocorrelation result $G(n)$ can be approximated as
\begin{equation}
G(n)\approx G_{\mathrm{y}}(n)+G_{\mathrm{i}}(n),
\label{eq:autocorrelation_components_approx}
\end{equation}

Similarly as in \cite{Schmidl_Cox_1997,Marey_2007_SC_narrowband}, two cases of the autocorrelation result (dependent on the value of $n$) can be considered:
%\begin{itemize}
%\item 
\subsubsection{Optimal timing point} 
In this case, repeatable samples of the preamble (distanced by $N/2$ sampling periods) are observed in the whole autocorrelation sliding-window at the receiver, so that $n=n_{\mathrm{opt}}$ where $n_{\mathrm{opt}}\in\{-N_{\mathrm{CP}}+L-1,...,0\}$. This is the Inter-Symbol Interference (ISI)-free and Inter-Carrier Interference (ICI)-free region of the time-sample index $n$, in which the repeatability of preamble samples is not distorted, i.e. $y(n_{\mathrm{opt}}+m)=y\left(n_{\mathrm{opt}}+m+\frac{N}{2}\right)$ for
$m\in\{0,...,N/2-1\}$. Having this in mind, the expected value of (\ref{eq:autocorrelation_components_G_x}) equals
\begin{equation}
\mathbb{E}\left[G_{\mathrm{y}}(n_{\mathrm{opt}})\right]\!=\!e^{\jmath \pi \nu}\!\!\!\sum_{m=0}^{N/2-1}\!\!\!\mathbb{E}\left[\left|y(n_{\mathrm{opt}}+m)\right|^{2}\right]\!=\!\sigma^{2}_{\mathrm{y}}\frac{N}{2}e^{\jmath \pi \nu},
\label{eq:autocorrelation_components_G_x_opt}
\end{equation}
assuming $y(n)$ has variance $\sigma^{2}_{\mathrm{y}}$ and zero mean\footnote{The same approach has been used e.g. in \cite{Marey_2007_SC_narrowband,Schmidl_Cox_1997}. Non-zero $\widetilde{x}(n)$ samples can be treated as a single realization of complex Gaussian random variable as shown in \cite{Wei_OFDM_Gaussian}, that allows for expectation calculation without any assumptions on $h(l)$ random values. Summation of N/2 consecutive components should keep the result for a single realization close to ensemble expectation.}.
The expected value of the function in the nominator of the S\&C timing metric is
\begin{equation}
\mathbb{E}\left[G(n_{\mathrm{opt}})\right]\!=\!\sigma^{2}_{\mathrm{y}}\frac{N}{2}e^{\jmath \pi \nu}+\frac{N}{2}\sigma^{2}_{\mathrm{i}}e^{\jmath \pi f}.
\label{eq:autocorrelation_components_opt}
\end{equation}
Depending on the received preamble power $\sigma^{2}_{\mathrm{y}}$, interference power $\sigma^{2}_{\mathrm{i}}$, CFO frequency $\nu$ and NBI frequency $f$, the timing metric can be degraded differently. Both complex components in (\ref{eq:autocorrelation_components_opt}), namely $\sigma^{2}_{\mathrm{y}}\frac{N}{2}e^{\jmath \pi \nu}$ and $\frac{N}{2}\sigma^{2}_{\mathrm{i}}e^{\jmath \pi f}$, can be represented as vectors on the complex-numbers plane. If the difference between $\nu$ and $f$ is close to zero (modulo 2), both vectors add up in phase and the S\&C algorithm performance is good. However, if both components have phases differing by $\pi$, i.e. normalized frequencies differ by one (modulo 2), the peak of $|G(n_{\mathrm{opt}})|$ will be maximally decreased, deteriorating the synchronization performance.
%\item 
\subsubsection{Outside the preamble} In this case, the autocorrelation window at the receiver does not include received preamble samples, so that the sampling moment equals $n=n_{\mathrm{out}}$, where $n_{\mathrm{out}}\in\{-\infty,...,-N_{\mathrm{CP}}-\frac{N}{2}\}\cup \{\frac{N}{2}+L-1,...,\infty\}$. In this case, the repeated samples of the NC-OFDM preamble are out of the correlation window and, as shown in \cite{Schmidl_Cox_1997}, the components of $G_{\mathrm{y}}(n_{\mathrm{out}})$ add up with random phases, so that, based on the Central Limit Theorem (CLT), $G_{\mathrm{y}}(n_{\mathrm{out}})$ is a complex Gaussian variable of a zero mean. Thus,
\begin{equation}
\mathbb{E}\left[G(n_{\mathrm{out}})\right]\!=\frac{N}{2}\sigma^{2}_{\mathrm{i}}e^{\jmath \pi f}.
\label{eq:autocorrelation_components_out}
\end{equation}
Note that in the absence of an NC-OFDM signal ($n\leq -N_{\mathrm{CP}}-\frac{N}{2}$), or when the NC-OFDM symbols other than the preamble are received ($n\geq \frac{N}{2}+L-1$), the false synchronization effect can occur, i.e., a peak in the synchronization metric can be observed because of the presence of NBI. The effect will be most significant in the first case ($n\leq -N_{\mathrm{CP}}-\frac{N}{2}$). As shown in \cite{Coulson_2004_synchr_narrow_sinusoid}, when there are no NC-OFDM signal samples over the autocorrelation window, the timing metric can achieve high values, and its expected value equals: 
\begin{equation}
\mathbb{E}\left[\left|\frac{G(n)}{M(n)}\right|^{2}\right]\approx \frac{\Gamma}{\Gamma+1}.
\label{eq:INR_vs_false}
\end{equation}
where $\Gamma$ is the interference to noise power ratio. 
%\end{itemize}

\section{NIRS Synchronization for NC-OFDM}
As shown above, the NC-OFDM spectrum nature (non-overlapping with the NBI spectrum) allows us to approximate 
 $G(n)$ by (\ref{eq:autocorrelation_components_approx}) for a relatively high SNR. The proposed NIRS algorithm aims at estimating component $G_{\mathrm{i}}(n)$, i.e., $\sigma_{\mathrm{i}}^{2}N/2e^{\jmath \pi f}$ and subtracting it from $G(n)$. It should remove the influence of NBI on $G(n)$ both for $n$ representing the optimum timing point and the time position outside the preamble period.  

Let us define the following function:
\begin{align}
&Q(n)=\frac{1}{2}\!\sum_{m=0}^{N/4-1}\!\Biggl[\!
r^{*}\!\left(\!m\!+\!n\right)\!r\!\left(\!m\!+\!n\!+\!\frac{N}{4}\!\right)\!+\!2r^{*}\!\left(\!m\!+\!n\!+\!\frac{N}{4}\!\right)\!
\nonumber \\&
\cdot \!r\!\left(\!m\!+\!n\!+\!\frac{N}{2}\!\right)\!
+\!r^{*}\!\left(\!m\!+\!n\!+\!\frac{N}{2}\!\right)\!r\!\left(\!m\!+\!n\!+\!\frac{3N}{4}\!\right)\!
\Biggr]
\label{eq:autocorrelation_NBI_est_improved}
\end{align}
that is based on the autocorrelation of the received samples with a constant delay of $\frac{N}{4}$ sampling periods. Assuming that the NC-OFDM system operates in an interference-limited environment, i.e., $\sigma^{2}_{\mathrm{i}}\gg\sigma^{2}_{\mathrm{w}}$, so that the noise power can be neglected ($w(n)\approx0$), the substitution of (\ref{eq:signal_model2}) into (\ref{eq:autocorrelation_NBI_est_improved}) results in
\begin{align}
&Q(n)\!\!=\!\!\frac{1}{2}\!\!\!\sum_{m=0}^{N/4-1}\!\!\Biggl[\!
\left(\!y^{*}(\!n\!+\!m\!)e^{\!-\jmath 2\pi \frac{\nu (n+m)}{N}}\!\!+\!\sqrt{\sigma^{2}_{\mathrm{i}}}e^{\!-\jmath 2\pi \frac{f (n+m)}{N}-\jmath\varphi}\right)
\nonumber \\ &
\cdot\left(y\left(n\!+\!m\!+\!\frac{N}{4}\!\right)e^{\jmath 2\pi \frac{\nu (n+m+\frac{N}{4})}{N}}\!+\!\sqrt{\sigma^{2}_{\mathrm{i}}}e^{\jmath 2\pi \frac{f (n+m+\frac{N}{4})}{N}+\jmath\varphi}\right)
\nonumber \\&
\!\!+\!\!2\!\left(\!\!y^{*}\!\!\left(\!n\!+\!m\!+\!\!\frac{N}{4}\!\right)\!e^{\!-\jmath 2\pi \frac{\nu (n+m+\frac{N}{4})}{N}}\!\!+\!\sqrt{\sigma^{2}_{\mathrm{i}}}e^{\!-\jmath 2\pi \frac{f (n+m+\frac{N}{4})}{N}-\jmath\varphi}\!\right)
\nonumber \\&
\cdot \left(\!y\left(\!n\!+\!m+\!\frac{N}{2}\!\right)e^{\!\jmath 2\pi \frac{\nu (n+m+\frac{N}{2})}{N}}\!\!+\!\sqrt{\sigma^{2}_{\mathrm{i}}}e^{\!\jmath 2\pi \frac{f (n+m+\frac{N}{2})}{N}+\jmath\varphi}\right) 
\nonumber \\&
\!\!+\!\!\left(\!y^{*}\!\!\left(\!n\!+\!m+\!\frac{N}{2}\!\right)e^{\!-\jmath 2\pi \frac{\nu (n+m+\frac{N}{2})}{N}}\!\!+\!\!\sqrt{\sigma^{2}_{\mathrm{i}}}e^{\!-\jmath 2\pi \frac{f (n+m+\frac{N}{2})}{N}-\jmath\varphi}\!\!\right)
\nonumber \\&
\cdot \!\!\left(\!y\!\left(\!n\!+\!m\!+\!\frac{3N}{4}\!\right)\!e^{\!\jmath 2\pi \frac{\nu (n+m+\frac{3N}{4})}{N}}\!\!+\!\sqrt{\sigma^{2}_{\mathrm{i}}}e^{\!\jmath 2\pi \frac{f (n+m+\frac{3N}{4})}{N}+\jmath\varphi}\right)\!\! 
\Biggr].
\label{eq:autocorrelation_NBI_est_improved1}
\end{align}
With a modicum of algebra, we obtain
\begin{align}
&Q(n)=Q_{\mathrm{i}}(n)+Q_{\mathrm{y}}(n)+Q_{\mathrm{cross}}(n),
\label{eq:autocorrelation_NBI_est_improved2}
\end{align}
where
\begin{align}
&Q_{\mathrm{i}}(n)=\frac{N}{2}\sigma^{2}_{\mathrm{i}}e^{\jmath \pi \frac{f}{2}},
\label{eq:autocorrelation_NBI_est_Q_I}
\end{align}
\begin{align}
&Q_{\mathrm{y}}(n)=\frac{e^{\jmath \pi \frac{\nu}{2}}}{2}
\sum_{m=0}^{\frac{N}{4}-1}\Biggl[
y^{*}(n+m)y\left(n+m+\frac{N}{4}\right)
\nonumber\\&
+2 y^{*}\left(n+m+\frac{N}{4}\right)y\left(n+m+\frac{N}{2}\right)
\nonumber\\&
+y^{*}\left(n+m+\frac{N}{2}\right)
y\left(n+m+\frac{3N}{4}\right)
\Biggr],
\label{eq:autocorrelation_NBI_est_Q_X}
\end{align}
\begin{align}
&Q_{\mathrm{cross}}(n)=\frac{\sqrt{\sigma^{2}_{\mathrm{i}}}}{2}e^{\jmath \pi \frac{f}{2}}\Biggl[
b^{*}(n)e^{\jmath \varphi}+b\left(n+\frac{N}{4}\right)e^{-\jmath \varphi}
\nonumber \\&
+b^{*}\left(n+\frac{N}{4}\right)e^{\jmath \varphi}+b\left(n+\frac{N}{2}\right)e^{-\jmath \varphi}
\Biggr].
\label{eq:autocorrelation_NBI_est_Q_CROSS}
\end{align}
In the above equations, $Q_{\mathrm{y}}(n)$ is the received NC-OFDM-signal-related term, $Q_{\mathrm{i}}(n)$ is the interference related term, and $Q_{\mathrm{cross}}(n)$ is the correlation between the NBI and the NC-OFDM signals. 
%Interestingly, similarly as in case of $G_{\mathrm{yi}}(n)$ and $G_{\mathrm{iy}}(n)$ defined by (\ref{Gyi}) and (\ref{Giy}), the component $Q_{\mathrm{cross}}(n)$ is a scaled sum of the DFTs of the incoming NC-OFDM signal on frequency $f-\nu$ (represented by $b(n)$ defined in (\ref{eq:DFT_z_cross})). 
As argued in Section \ref{sec:characteristic_S_C_NBI} and formally derived in the Appendix, the values of $b(n)$ are negligible in the case of NC-OFDM transmission (as opposed to the standard OFDM) and as such $Q_{\mathrm{cross}}(n)$ can be omitted in (\ref{eq:autocorrelation_NBI_est_improved2}). As a result, formula (\ref{eq:autocorrelation_NBI_est_improved2}) can be approximated as:
\begin{align}
&Q(n)\approx Q_{\mathrm{i}}(n)+Q_{\mathrm{y}}(n),
\label{eq:autocorrelation_NBI_est_approx}
\end{align}  
As in Section \ref{sec:characteristic_S_C_NBI}, for $G(n)$, two cases of $n$ value can be considered:
%\begin{itemize}
%\item 
\setcounter{subsubsection}{0}
\subsubsection{Optimal timing point} In this case, again, the autocorrelation window at the receiver contains all repeatable samples (distanced by $N/2$ sampling periods) of the preamble, so that $n=n_{\mathrm{opt}}$. There is no ICI or ISI and $y(n_{\mathrm{opt}}+m)=y\left(n_{\mathrm{opt}}+m+\frac{N}{2}\right)$ for
$m\in\{0,...,N/2-1\}$. Having this in mind, $Q_{\mathrm{y}}(n_{\mathrm{opt}})$ equals:
\begin{align}
&Q_{\mathrm{y}}(n_{\mathrm{opt}})=\frac{e^{\jmath \pi \frac{\nu}{2}}}{2}
\sum_{m=0}^{\frac{N}{4}-1}\Biggl(
2y^{*}(\!n_{\mathrm{opt}}\!+\!m\!)y\!\left(\!n_{\mathrm{opt}}\!+\!m\!+\!\frac{N}{4}\!\right)
\nonumber \\ &
+2y(n_{\mathrm{opt}}+m)y^{*}\left(n_{\mathrm{opt}}+m+\frac{N}{4}\right)
\Biggr)\label{eq:autocorrelation_NBI_Q_X_opt}
% \nonumber
 \\ &
=2e^{\jmath \pi \frac{\nu}{2}}
\sum_{m=0}^{\frac{N}{4}-1}
\Re\left(y^{*}(n_{\mathrm{opt}}+m)y\left(n_{\mathrm{opt}}+m+\frac{N}{4}\right)\right)
\nonumber
,
\end{align}  
where $\Re(\cdot)$ denotes the real part of a complex number. Because the preamble samples distanced by $\frac{N}{4}$ sampling periods are uncorrelated, $\frac{N}{4}$ samples add up with random phases\cite{Schmidl_Cox_1997}, so that, based on the CLT, $Q_{\mathrm{y}}(n_{\mathrm{opt}})$ is the zero-mean Gaussian random variable. The expected value of $Q(n_{\mathrm{opt}})$ equals then
\begin{align}
&\mathbb{E}\left[Q(n_{\mathrm{opt}})\right]=\frac{N}{2}\sigma^{2}_{\mathrm{i}}e^{\jmath \pi \frac{f}{2}}
\label{eq:autocorrelation_NBI_est_opt}
\end{align}     
%\item 
\subsubsection{Outside the preamble} In this case, again, there are no repeated preamble sample pairs (distanced by $N/2$ sampling periods) within the receiver autocorrelation window, so that the sampling moment equals $n=n_{\mathrm{out}}$, for which, similarly as in the optimum timing point, NC-OFDM symbol samples distanced by $\frac{N}{4}$ sampling periods are uncorrelated, i.e., each product in (\ref{eq:autocorrelation_NBI_est_Q_X}) has a zero expected value, giving $\mathbb{E}\left[Q_{\mathrm{y}}(n_{\mathrm{out}})\right]=0$. This means that the expected value of $\mathbb{E}\left[Q(n_{\mathrm{out}})\right]$ is the same as in the optimum timing point, i.e.,
\begin{align}
&\mathbb{E}\left[Q(n_{\mathrm{out}})\right]=\frac{N}{2}\sigma^{2}_{\mathrm{i}}e^{\jmath \pi \frac{f}{2}}
\label{eq:autocorrelation_NBI_est_out}
\end{align}  
%\end{itemize}

Because the argument of the complex number $\mathbb{E}\left[Q(n)\right]$ (both for $n=n_{\mathrm{opt}}$ and $n=n_{\mathrm{out}}$) is a half of the argument in $G_{\mathrm{i}}(n)$, it has to be doubled to remove the NBI impact from  (\ref{eq:autocorrelation_components_approx}).
%(\ref{eq:autocorrelation_result_approximation}). 
The proposed NIRS timing metric nominator $G_{\mathrm{NIRS}}(n)$ is thus:
\begin{equation}
G_{\mathrm{NIRS}}(n)=G(n)-|Q(n)|e^{\jmath 2\arg\{Q(n)\}}.
\label{eq:improved_second_estimator}
\end{equation}
The decision on the beginning of the frame in time is made for $n$ maximizing the decision metric $|G_{\mathrm{NIRS}}(n)/M(n)|^{2}$:  
\begin{equation}
\widehat{n}= \argmax_{n} \left|\frac{G_{\mathrm{NIRS}}(n)}
{M(n)}\right|^{2}\mathrm{.}
\label{eq:decision_metric_proposed}
\end{equation}  
Moreover, the sampling moment in the middle between two sampling moments for which 90\% of the maximum of $|G_{\mathrm{NIRS}}(n)/M(n)|^{2}$ is obtained, can be used for time-synchronization decision (similarly as proposed by Schmidl \& Cox). 

\begin{figure}[t!]
\centering
\includegraphics[width=3.4in]{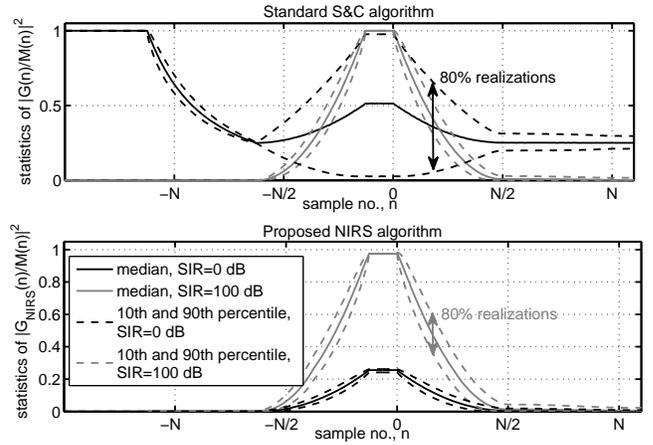}
\caption{Statistics of S\&C and NIRS timing metrics: median (solid line) and 10/90th percentile (dashed line). NC-OFDM system not distorted (SIR=100dB) and distorted by strong NBI (SIR=0dB) and random CFO.}
\label{fig_example_of_both_autocorrelations}
\end{figure}

The examples of timing-metric statistics are presented in Fig. \ref{fig_example_of_both_autocorrelations} for the S\&C and NIRS algorithms.  The example system parameters are: $N=256$, $N_{\mathrm{CP}}=N/8$, $L=1$, $h(0)=1$, random CFO drawn from the uniform distribution in the range of (--10.5 kHz; 10.5 kHz) as used in Sec. \ref{sec:Simulations},
and SNR$=\infty$. Statistics have been measured over $10^4$ NC-OFDM frames, each consisting of $P=11$ modulated, random NC-OFDM symbols and preceded by 3 empty symbols. The timing metrics are presented in grey in the case of negligible NBI (SIR~=~100~dB\footnote{ SIR is defined as the signal-power to the interference-power ratio over the whole NC-OFDM receiver band, and is calculated over the time when non-zero NC-OFDM symbols are present (i.e. empty NC-OFDM symbols before a frame are excluded from calculations). Similar assumptions are made for the SNR definition.}). In this case, as expected, the S\&C method achieves its maximum of $1$ for $n=n_{\mathrm{opt}}$ that is around $n=0$. Moreover, the timing metric value is close to zero for $n$ being outside the preamble. This result is independent from the symbols modulating the NC-OFDM preamble or the CFO value, because the timing metric variance\footnote{ Observe that the variance is directly related to the difference between 90th and 10th percentile visible in Fig. \ref{fig_example_of_both_autocorrelations}.} is small and the median is close to zero. Interestingly, the statistics for the NIRS method are very similar to those of the S\&C metric. 
The black lines denote a set-up when the NC-OFDM signal is distorted by the NBI modeled as a complex sinusoid and SIR~=~0~dB. The S\&C algorithm has a plateau of the timing metric equal approximately to 1 before the preamble begins, because a high value of (\ref{eq:INR_vs_false}) is obtained as $\Gamma=\infty$. It causes a false synchronization, because the timing metric has a maximum before the NC-OFDM frame begins. Additionally, the S\&C timing metric has a high variance around the optimal timing point. It is caused by varying phases in the $G(n_{\mathrm{opt}})$ components, as it has been shown in (\ref{eq:autocorrelation_components_opt}).
% (by varying difference between $\nu$ and $f$). 
As expected, the NIRS algorithm succeeds to estimate and remove the NBI effect and is robust against the false synchronization phenomenon. It is visible in the low median and variance values of the timing metric before preamble begins. For the optimal timing point, the NIRS algorithm achieves a lower peak than in the case of no-NBI. It is caused by $M(n)$ value that composes of the energy of the usefull NC-OFDM signal, white noise and the NBI. However, the variance of the proposed estimator is low, so correct peak detection is highly probable.

After finding the synchronization in time, the CFO can be estimated as 
\begin{equation}
\widehat{\nu}=\arg \left\{\frac{G_{\mathrm{NIRS}}(\widehat{n})}{\pi}\right\}
\label{eq:decision_metric_second_imporoved_frac_frequency}.
\end{equation}   

Importantly, the proposed estimator is not only working correctly for the high NBI power. It also has no variance in the no-noise, no-interference scenario assuming correct time synchronization. In this case, (\ref{eq:autocorrelation_NBI_est_improved2}) can be considered for $Q_{\mathrm{cross}}(n_{\mathrm{opt}})=Q_{\mathrm{i}}(n_{\mathrm{opt}})=0$, i.e. $Q(n_\mathrm{opt})= Q_{\mathrm{y}}(n_{\mathrm{opt}})$. 
Although the expected value of this component equals zero, it could have non-zero variance. Component (\ref{eq:autocorrelation_NBI_Q_X_opt}) can be introduced into (\ref{eq:improved_second_estimator}) for $Q(n)$. Interestingly, the complex number argument of $Q(n_\mathrm{opt})$ to be doubled is  $\pi \frac{\nu}{2}$ or $\pi \frac{\nu}{2}+\pi$, depending on the summation result in (\ref{eq:autocorrelation_NBI_Q_X_opt}) being positive or negative, respectively. However, in both cases, the doubled complex number argument (as done in (\ref{eq:improved_second_estimator})) is $\pi \nu$ (modulo $2\pi$). The nominator of the NIRS algorithm metric can be rewritten using  (\ref{eq:autocorrelation_components_G_x}) for the optimal timing point and (\ref{eq:autocorrelation_NBI_Q_X_opt}):
\begin{align}
&G_{\mathrm{NIRS}}(n_\mathrm{opt})=e^{\jmath \pi \nu}
\sum_{m=0}^{N/2-1}
\left|y(n_{\mathrm{opt}}+m)\right|^{2}
\\&\nonumber
-2e^{\jmath \pi \nu}
\Bigl|\sum_{m=0}^{\frac{N}{4}-1}
\Re\left(y^{*}(n_{\mathrm{opt}}+m)y\left(n_{\mathrm{opt}}+m+\frac{N}{4}\right)\right)\Bigr|.
\end{align}
Both summations in the above equation result in real positive values (because of the absolute-value operation) and have a common complex coefficient $e^{\jmath \pi \nu}$, so that the application of formula (\ref{eq:decision_metric_second_imporoved_frac_frequency}) should result in the perfect CFO estimate. However, if the second sum has a higher absolute value than the first one, the resultant complex number argument will equal $\pi \nu +\pi$, i.e., a high CFO estimation error will occur. The correctness of the proposed metric in the considered environment can be proved by replacing the second absolute value operator by $\pm$ (dependent on the summation under the absolute value having a positive or negative result) and decomposing each complex number into real and imaginary parts, i.e., 
\begin{align}
&\!\!\!G_{\mathrm{NIRS}}\!(n_\mathrm{opt}\!)\!\!=\!\!e^{\jmath \pi \nu}\!\!\!
\sum_{m=0}^{N/4-1}\!\!
\Biggl(\!\Re\left(y(n_{\mathrm{opt}}\!+\!m)\right)^{2}\!\!+\!\Im\!\left(y(n_{\mathrm{opt}}\!+\!m)\right)^{2}
\nonumber% \\ &
\end{align}
\begin{align}
&
+\Re\left(y\!\left(n_{\mathrm{opt}}\!+\!m\!+\!\frac{N}{4}\right)\right)^{2}\!
%\nonumber \\ &
+\Im\left(y\!\left(n_{\mathrm{opt}}\!+\!m\!+\!\frac{N}{4}\right)\right)^{2}
\Biggr) 
\nonumber \\&\nonumber
\pm 2e^{\jmath \pi \nu}
\sum_{m=0}^{\frac{N}{4}-1}\Biggl(
\Re\left(y(n_{\mathrm{opt}}\!+\!m)\right)\Re\left(y\left(n_{\mathrm{opt}}\!+\!m\!+\!\frac{N}{4}\right)\right)
\\&
%\nonumber
+\Im\left(y(n_{\mathrm{opt}}+m)\right)\Im\left(y\left(n_{\mathrm{opt}}+m+\frac{N}{4}\right)\right)
\Biggr)
\nonumber \\ &
=\!e^{\jmath \pi \nu}\!\! \sum_{m=0}^{N/4-1}\!\!\Biggl(\!
%\nonumber \\ &
\left(\!\Re\left(y(n_{\mathrm{opt}}\!+\!m)\!\right)\!\pm \Re\!\left(\!y\!\left(\!n_{\mathrm{opt}}\!+\!m\!+\!\frac{N}{4}\right)\!\right)\!\right)^{2}
\nonumber \\ &
+
\left(\Im\left(y(n_{\mathrm{opt}}\!+\!m)\right)\pm \Im\left(y\left(n_{\mathrm{opt}}\!+\!m\!+\!\frac{N}{4}\right)\right)\right)^{2}
\Biggr),
%\nonumber
\end{align}    
where $\Im(\cdot)$ denotes the imaginary part of a complex number. The above summation cannot have a negative result and as such, it confirms that the above estimator has a zero variance in a no-noise, no-interference scenario.

\section{Computational Complexity}
An important advantage of the S\&C algorithm is its low computational complexity. Importantly, both functions in the nominator and in the denominator of the timing metric can be calculated iteratively, i.e.,
\begin{align}
G(n)=&G(n-1)-r^{*}(n-1)r\left(n-1+\frac{N}{2} \right)
\nonumber \\ &
+r^{*}\left(n+\frac{N}{2}-1\right)r\left(n-1+N\right),
\end{align}
\begin{align}
&M(n)\!=\!M(n\!-\!1)\!-\!\left|r\!\left(n\!+\!\frac{N}{2}\!-\!1\!\right)\right|^{2}
\!\!+\!\left|r\left(n\!+\!N\!-\!1\right)\right|^{2}.
\end{align}
 The computational complexity of the S\&C algorithm can be estimated by a number of arithmetic operations needed to calculate the timing metric $|G(n)/M(n)|^{2}$ for a single $n$. The number of real additions/subtractions and real multiplications/divisions for this algorithm is provided in Table \ref{table_computational_compl}.
\begin{table}[th]
\caption{ Operations count per single input sample $n$ for $N=256$}
\label{table_computational_compl}
\begin{center}
\begin{tabular}{p{1.1cm}ll}
\hline
Algorithm & Real additions/subtractions & Real multiplications/divisions \\ 
\hline
S\&C & $10$ & $10$ \\ 
Ziabari & $16N+11=4107$ & $16N+22=4118$ \\ 
AHD1 (iterative) & $12N+49=3121$ & $10\frac{3}{4}N+65=2817$ \\
NIRS & $10+10+4=24$ & $10+6+8=24$ \\
% & $3N\log_{2}N=6902$ & $2N\log_{2}N=5448$\\
\hline
\end{tabular}
\end{center}
\end{table}     

The NIRS algorithm requires the same number of operations as the S\&C algorithm, but additionally, $Q(n)$ has to be calculated and subtracted from $G(n)$ as shown in (\ref{eq:improved_second_estimator}). Importantly, $Q(n)$ can be calculated iteratively 
 \begin{align}
&Q(n)=Q(n-1)+\frac{1}{2}\Biggl(
r^{*}\left(n+\frac{3N}{4}-1\right)r\left(n-1+N\right)
\nonumber \\ &
-r^{*}\left(n-1\right)r\left(n-1+\frac{N}{4}\right)
%\nonumber \\ &
+r^{*}\left(n+\frac{N}{2}-1\right)
%\nonumber
 \\ & \cdot
 r\left(n-1+\frac{3N}{4}\right)
%\nonumber \\ &
-r^{*}\left(n-1+\frac{N}{4}\right)r\left(n-1+\frac{N}{2}\right)
\Biggr),
\nonumber
\end{align}
that needs only one complex multiplication in each iteration, i.e., $r^{*}\left(n+\frac{3N}{4}-1\right)r\left(n-1+N\right)$.
%, because the other products have been already calculated. 
Therefore, the calculation of a single $Q(n)$ value requires $10$ real additions/subtractions and $6$ real multiplications. Another problem is the efficient doubling of the $Q(n)$ argument. Note that
\begin{align}
|Q(n)|e^{\jmath 2\arg\{Q(n)\}}=\frac{Q(n)^{2}}{|Q(n)|}
\label{eq:computation_of_doubled_Q_n}
\end{align}
requires $4$ real multiplications and $1$ real addition to calculate the nominator. The complex number modulus (in the denominator of (\ref{eq:computation_of_doubled_Q_n})) can be calculated with $2$ real multiplications, $1$ addition and one square root operation (omitted in Table \ref{table_computational_compl}). In practice, it is an iterative CORDIC (Coordinate Rotation Digital Computer) \cite{Volder_CORDIC_1959} algorithm that is used for this purpose. Because the amplitude and phase of $Q(n)$ and $Q(n-1)$ should be close to each other, the number of CORDIC iterations can be significantly reduced. The last operations needed for $G_{\mathrm{NIRS}}(n)$ calculation are two real divisions and one complex subtraction. The total NIRS computational complexity (including the basic S\&C complexity) is shown in the last row of Table \ref{table_computational_compl}. 

For comparison, the computational complexity of the timing-metric calculation of Ziabari's \cite{Ziabari_synchr_NC_OFDM} and the \emph{Ad-Hoc Detector no. 1} (AHD1) \cite{Sanguinetti_2010_synchr} synchronization algorithms are also shown in Table \ref{table_computational_compl}. (These algorithms have been examined in the simulation experiments for comparison purposes --see Section \ref{sec:Simulations}.) This complexity has been assessed for Ziabari's cross-correlation-like metric of $4N$ components. 

It is visible that the proposed NIRS algorithm has a computational complexity more than twice as high as the one of the S\&C algorithm. However, in comparison to Ziabari's and AHD1 methods (both designed as robust against the false-synchronization effect), the computational complexity of NIRS is nearly negligible.

\section{Simulation Results}
\label{sec:Simulations}
The proposed NIRS algorithm has been evaluated against the standard S\&C \cite{Schmidl_Cox_1997}, NBI-robust AHD1 algorithm \cite{Sanguinetti_2010_synchr}, and cross-correlation-based Ziabari's method \cite{Ziabari_synchr_NC_OFDM}. 

The simulation set-up is the following. A frame consists of a preamble and 10 QPSK-mapped NC-OFDM symbols ($P=11$), and is preceded by 3 empty symbols emulating bursty transmission, $N=256$, and $N_{\mathrm{CP}}=N/8$. In each examined case, $10^{5}$ frames of modulated random symbols have been transmitted. Although the symbols modulating SCs in the preamble should be usually deterministic, e.g., generated using Zadoff-Chu sequence as in the case of LTE, it is a common approach to test algorithms using random symbols (as in \cite{Schmidl_Cox_1997} for the conventional S\&C algorithm). Moreover, simulations not presented here confirm that the same performance relations are observed for different synchronization algorithms for both Zadoff-Chu and random preambles. Selection of an optimal preamble in the case of NC-OFDM is much more complicated than in the OFDM system, as the utilized band is not continuous. The SC spacing is 15~kHz as in the LTE system. The channel model is the COST 207 6-path Typical Urban Rayleigh fading channel. The set of data SC is $\mathbf{I}=\{-100,...,-1,1,...,3,46,...,100\}$. The Ziabari method uses $4N$ components in its correlation metric. In the AHD1 algorithm, $\lambda$ equal to 0.1 is used as in \cite{Sanguinetti_2010_synchr}. The NBI normalized frequency is $24.5$, and is modeled as an "ideal" NBI, i.e., 
unmodulated complex sinusoid, or in another case, as a "practical" NBI, i.e., an FM modulated signal of a 200 kHz bandwidth (e.g., wireless microphone --WM) modeled as in \cite{Fuhrwerk_FBMC_microph}. The spectral notch of 42 subcarriers is enough to accommodate WM transmission allowing for the high NC-OFDM system throughput and the acceptable interference power in the WM receiver according to the calculations in \cite{KryszkiewiczEW12}. Each NC-OFDM frame is distorted by a random CFO of uniform distribution over $(-10.5$ kHz$, 10.5 $kHz$)$, chosen arbitrarily. In any case of the NBI, it is distorted by a random CFO having uniform distribution over $(-14$ kHz$, 14~$kHz$)$. (According to \cite{ETSI_mikrofon_jak_mierzyc_maske}, the frequency instability in the WM case can be high, i.e., 20 ppm, that gives 14 kHz at carrier frequency 700 MHz mentioned in the introduction.)
For a fair comparison, the Ziabari's method CFO search range was limited to be similar as in the NIRS and S\&C methods. 
\begin{figure}[!t]
\centering
\includegraphics[width=3.4in]{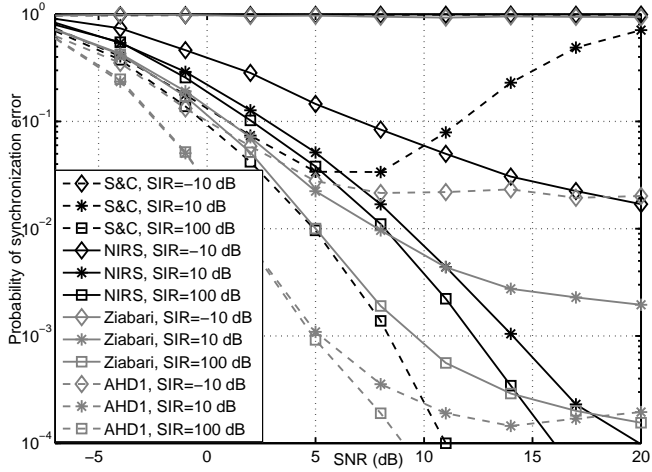}
\caption{Probability of synchronization error for the case of "practical" NBI.}
\label{fig_prob_real_interf}
\end{figure}
\begin{figure}[!t]
\centering
\includegraphics[width=3.4in]{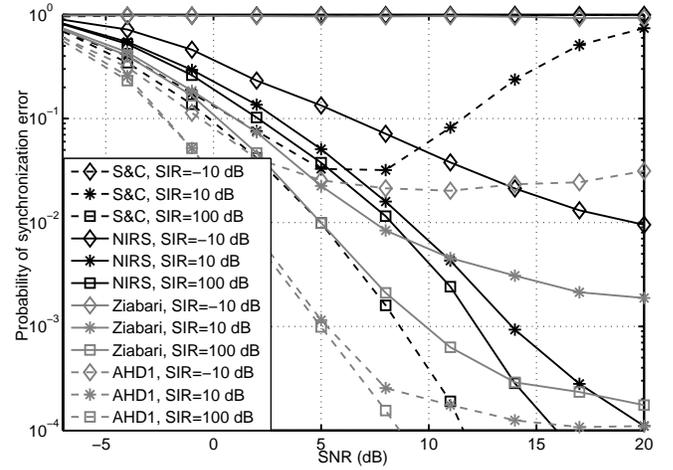}
\caption{Probability of synchronization error for the case of "ideal" NBI.}
\label{fig_prob_ideal_interf}
\end{figure}

In Fig. \ref{fig_prob_real_interf} and Fig. \ref{fig_prob_ideal_interf}, the probabilities of a synchronization error are shown in the case of the "practical" and "ideal" NBI, respectively. The frame is treated as erroneously synchronized 
if an error in the time domain is higher than the CP length, i.e., 
when  $|\widehat{n}|>N_{\mathrm{CP}}$, or if an error in the CFO estimate is higher than 0.5 of SC spacing. Although it is known that for high-order constellations, more precise synchronization is required, this metric provides a good estimate of the gross error of synchronization algorithms.   
Note that the NIRS algorithm outperforms all other algorithms in high-SNR and low-SIR scenarios (for which it was designed). Under these conditions, the S\&C algorithm is also prone to false synchronization. 
It is visible as high probability of the synchronization error for SIR$\leq 10$~dB and high SNR.
For practical SNR values, e.g., SNR$> 10~$dB, our NIRS algorithm has a higher probability of synchronization error than S\&C only for SIR$=100$~dB \footnote{ Observe that SIR$=100$~dB approximates the scenario without the NBI presence.}. It is caused by the non-zero variance of estimator $Q(n)$ that increases the variance of the timing metric in comparison to the S\&C method. Still, the SNR required to correctly synchronize at least $99\%$ of frames equals 8~dB, that should be acceptable. It is an increase of about 3~dB in comparison to the S\&C algorithm. Although the AHD1 algorithm has a better performance for a low SNR, this is obtained at the cost of a much higher computational complexity. Additionally, the required preamble consists of 4 identical sequences (with one sign inversion) that may increase interference to the PU-band.
The NBI model has a negligible influence on the S\&C and Ziabari methods. Interestingly, the FM modulated signal of a 200~kHz bandwidth ("practical" NBI) is still narrowband enough to cause the false synchronization effect in the S\&C algorithm. The performance of NIRS depends on the NBI model only for the highest NBI power, i.e., SIR=-10~dB. The inaccuracy of "practical" NBI modeling in the NIRS algorithm by (\ref{eq:signal_model2}) requires an SNR higher by 0.6~dB for the probability of synchronization error equal to $0.1$ in comparison to an "ideal" interference case (for which (\ref{eq:signal_model2}) perfectly models the NBI). However, even in the case of an "ideal" NBI, the NIRS synchronization performance is degraded for a high-power NBI. This is caused by the non-zero variance of $Q_{\mathrm{cross}}(n)$ and $G_{\mathrm{cross}}(n)$, though significantly reduced in the NC-OFDM case by near-zero $b(n)$ values, when compared to the standard OFDM as shown in Appendix \ref{sec:appendix}.        

\begin{figure}[!t]
\centering
\includegraphics[width=3.4in]{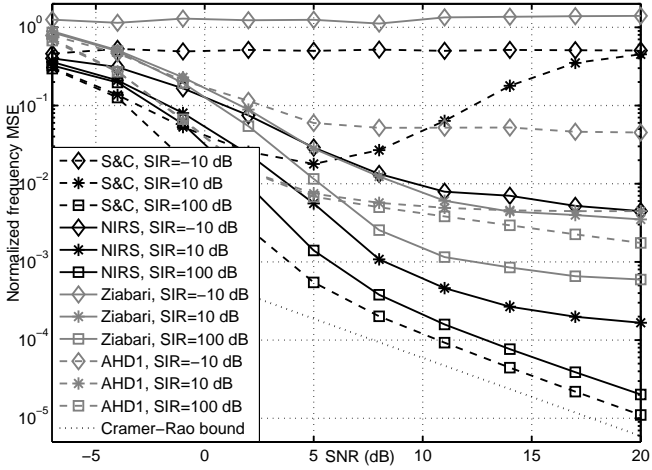}
\caption{MSE of frequency estimate for the case of "practical" NBI.}
\label{fig_MSE_real_interf_freq}
\end{figure}

\begin{figure}[!t]
\centering
\includegraphics[width=3.4in]{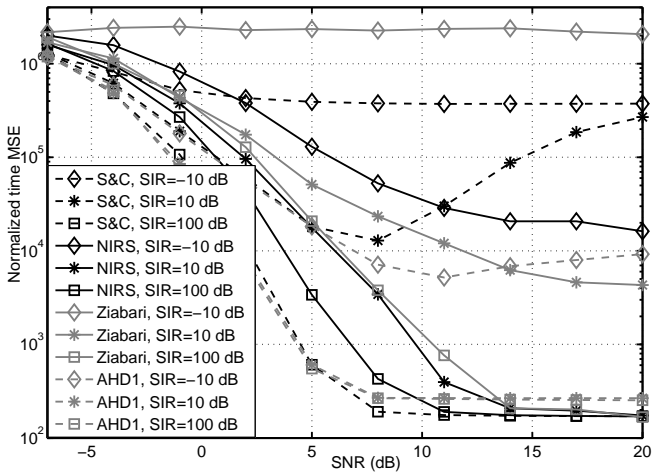}
\caption{MSE of time estimate for the case of "practical" NBI.}
\label{fig_MSE_real_interf_time}
\end{figure}

\begin{figure}[!t]
\centering
\includegraphics[width=3.4in]{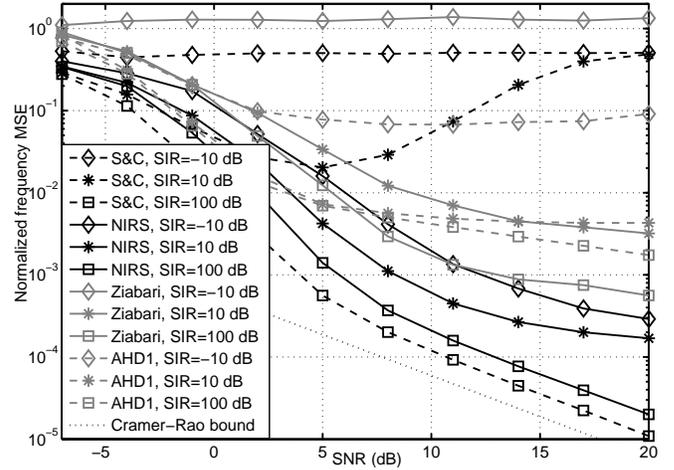}
\caption{MSE of frequency estimate for the case of "ideal" NBI.}
\label{fig_MSE_ideal_interf_freq}
\end{figure}

\begin{figure}[!t]
\centering
\includegraphics[width=3.4in]{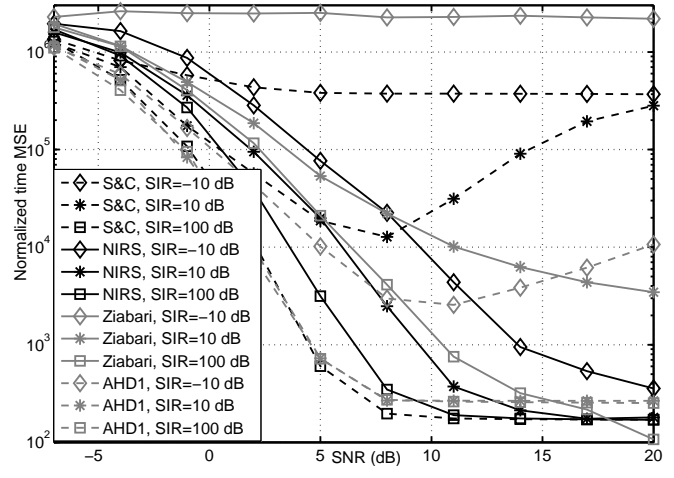}
\caption{MSE of time estimate for the case of "ideal"' NBI.}
\label{fig_MSE_ideal_interf_time}
\end{figure}

The MSEs of frequency (normalized to SC spacing) and time (normalized to the sampling period) offsets are shown for the "practical" NBI model in Fig. \ref{fig_MSE_real_interf_freq} and Fig. \ref{fig_MSE_real_interf_time}, respectively. Similarly, in the case of "ideal" NBI, MSEs are shown in Fig. \ref{fig_MSE_ideal_interf_freq} and Fig. \ref{fig_MSE_ideal_interf_time}. Similarly as visible on the previous plots, false synchronization occurs in the S\&C method, for high-SNR, and low-SIR scenarios, which is visible both for the time and frequency MSEs. Interestingly, for SIR~=~100~dB, NIRS requires only about 2~dB of SNR increase in comparison to the S\&C algorithm for a time MSE equal to $10^{3}$ and a frequency MSE equal to $10^{-3}$. This means that the frame erroneously synchronized by the NIRS algorithm typically has the time and frequency error relatively small. The opposite effect occurs in the case of Ziabari's method. A small number of incorrectly synchronized frames has a high error in time and frequency, causing high MSE values. Most importantly, it is visible that NIRS achieves lower time and frequency MSEs when NBI is practically present (SIR$<100~$dB) in the received signal in comparison to the S\&C and Ziabari's algorithms for the case of practical white noise power, e.g., SNR$>5~$dB. Observe that even for nearly no-noise (SNR=20 dB) and no-interference (SIR=100 dB) scenario, AHD1 obtains relatively high frequency MSE, i.e., higher than $10^{-3}$.
    

 
\begin{figure}[!t]
\centering
\includegraphics[width=3.4in]{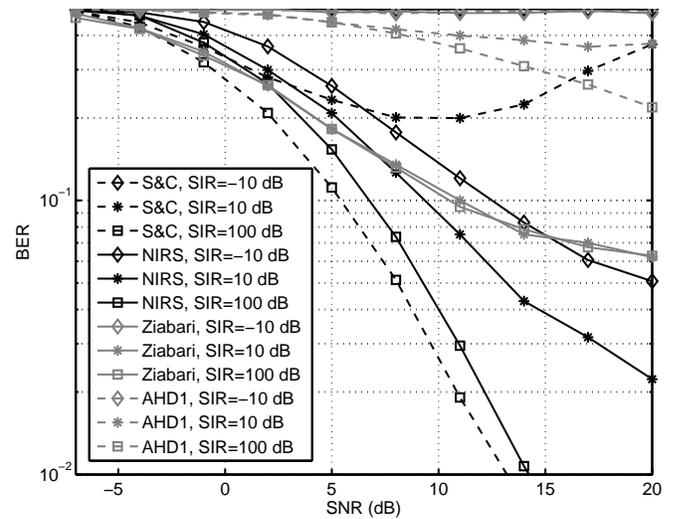}
\caption{BER of preamble QPSK symbols detection for the case of "ideal" NBI.}
\label{fig_BER_ideal}
\end{figure}
The ultimate goal of the synchronization algorithm is to allow reliable bits reception. For all considered algorithms, the Bit Error Rates (BERs) of the reception of QPSK preamble symbols have been calculated for various SIR and SNR values as shown in Fig.\ref{fig_BER_ideal}. For these simulations, the ideal NBI model has been used while other parameters have been adopted the same as used to obtain the results presented in Fig. \ref{fig_prob_ideal_interf}. While time and frequency offsets have been estimated according to a given synchronization algorithm, perfect channel knowledge and Zero-Forcing equalization have been assumed. 

Note that although NIRS obtains SNR loss of about 1 dB in comparison to S\&C algorithm for SIR=100 dB, it outperforms other algorithms in this scenario. Under severe interference (SIR~=~--10 dB), it is the only algorithm obtaining BER$<10^{-1}$. Even though in many cases, the AHD1 algorithm achieves lower probability of synchronization error, it has higher frequency MSE floor, resulting in high BER. AHD1 would require the second detection stage for the CFO estimation improvement. The results in the case of "practical" NBI lead to the same conclusions.       

\begin{figure}[!t]
\centering
\includegraphics[width=3.4in]{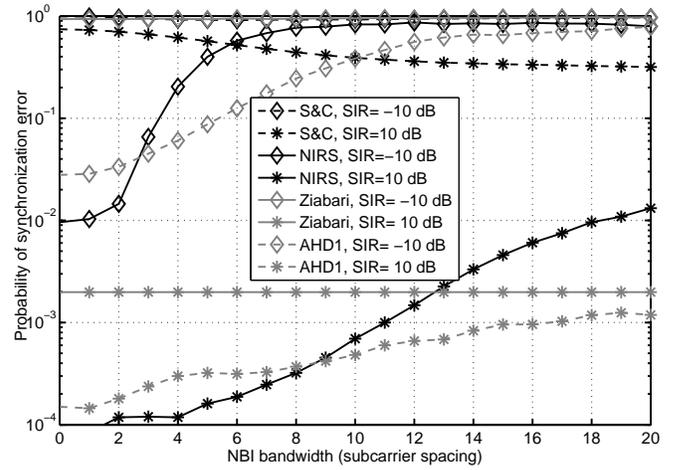}
\caption{Probability of synchronization error for varying NBI bandwidth.}
\label{fig_Prob_changed_NBI_bandwidth}
\end{figure}

%\subsection{Computational complexity}
Additional computer simulations have been carried out in order to estimate the influence of the NBI bandwidth on synchronization performance. The simulation set-up is the same as previously, except for the NBI model. Now, it is a frequency-modulated complex sinusoid using a 1 kHz sinusoid signal of a random initial phase. The frequency deviation is adjusted in order to obtain the required NBI bandwidth (calculated using Carson's rule). While 3 different SIR values are used, a constant SNR of 20 dB is assumed. In Fig. \ref{fig_Prob_changed_NBI_bandwidth}, the probability of a synchronization error is presented.
It is visible that NIRS performs the better, the narrower the NBI bandwidth is, and as the signal model in (\ref{eq:signal_model2}) becomes more accurate. On the other hand, the S\&C algorithm in most cases fails to synchronize, although it is visible that the wider the NBI bandwidth, the lower the obtained probability of the synchronization error. It is caused by the reduced probability of "false synchronization", as the interference has lower autocorrelation values. It is visible that while Ziabari's method performance is independent from the interference bandwidth, AHD1 performs the better, the narrower the NBI bandwidth. However, for a strictly narrowband interference, the proposed NIRS algorithm outperforms AHD1, with a much lower computational complexity required. 

%MAM TEZ WYKRESY MSE ALE TO CHYBA NADMIAROWE
%Similar conclusions can be drawn from MSE plots shown in Fig........   

\section{Conclusion}
The NIRS algorithm proposed in this paper for an NC-OFDM- based CR system has a good  performance in the presence of a severe licensed-user-originated NBI. Importantly, this is obtained at a very low increase of computational complexity. It is therefore suitable for low-power, battery-operated NC-OFDM-based cognitive radio terminals.

% or
\appendix[Justification of a low $b(n)$ value in NC-OFDM]
\label{sec:appendix}
The aim is to prove that $b(n)$ is low in the case of the NC-OFDM signal in comparison to the standard OFDM case. Through the substitution of (\ref{eq:channel_influence}) to (\ref{eq:DFT_z_cross}), we obtain:
\begin{equation}
b(n)=\sum_{l=0}^{L-1}h(l)
\sum_{m=0}^{\frac{N}{2}-1}\tilde{x}(n+m-l)e^{-\jmath 2 \pi (f-\nu) \frac{n+m}{N}}.
\label{eq:appendix1}
\end{equation}
Note that all signal components coming from different channel paths can be calculated separately and added up. Additionally, the summation window of $\frac{N}{2}$ samples (over $m$) can span a maximum of 2 consecutive NC-OFDM symbols. Because of additivity, the influence of each symbol on $b(n)$ can be calculated separately and added afterwards, i.e.,
\begin{equation}
b(n)=\sum_{l=0}^{L-1}h(l)\left(b^{(p)}_{n-l}+b^{(p+1)}_{n-l}\right),
\label{eq:appendix2}
\end{equation}
\begin{equation}
b^{(p)}_{n}=\sum_{m=0}^{\frac{N}{2}-1}x_{n+m}^{(p)}e^{-\jmath 2 \pi (f-\nu) \frac{n+m}{N}}
\label{eq:appendix3}
\end{equation}
where the $p$-th and $p+1$-th NC-OFDM symbols have to be chosen according to $n$. 
Without the loss of generality, the calculations are done for $p=0$. The $b^{(0)}_{n}$ solution is found by substituting (\ref{eq_IFFT}) into (\ref{eq:appendix3}), and by limiting the range of $m$, so that $n+m\in\{-N_{\mathrm{CP}},...,N-1\}$ (which is the range of non-zero values of $x_{n+m}^{(0)}$). There are 3 possible cases:
\paragraph{The calculation window spans partially over the beginning of the $0$-th NC-OFDM symbol, i.e., $n\in\{-\frac{N}{2}-N_{\mathrm{CP}}+1,...,-N_{\mathrm{CP}}-1\}$, giving} 
\begin{align}
\label{eq:appendix4}
&b^{(0)}_{n}\!=\!\!\!\!\!\sum_{m=-N_{\mathrm{CP}}-n}^{\frac{N}{2}-1}
\sum_{k=-N/2}^{N/2-1}\!\!
\frac{d_{k}^{(0)}}{\sqrt{N}}
e^{\jmath 2\pi \frac{(n+m)k}{N}}e^{-\jmath 2 \pi (f-\nu) \frac{n+m}{N}}
 \\ &
=\sum_{k=-N/2}^{N/2-1}\frac{d_{k}^{(0)}}{\sqrt{N}}e^{\jmath 2\pi \frac{n}{N}(k-f+\nu)}
\sum_{m=-N_{\mathrm{CP}}-n}^{\frac{N}{2}-1} e^{\jmath 2\pi \frac{m}{N}(k-f+\nu)} 
\nonumber 
\\ &
=\!\!\sum_{k=-N/2}^{N/2-1}\!\!\frac{d_{k}^{(0)}}{\sqrt{N}}e^{\jmath 2\pi \frac{n}{N}(k-f+\nu)}
\frac{
e^{\jmath 2\pi \frac{-N_{\mathrm{CP}}-n}{N}(\!k\!-\!f\!+\!\nu\!)}
\!-\!
e^{\jmath \pi (\!k\!-\!f\!+\!\nu\!)}
}{
1-e^{\jmath 2\pi \frac{k-f+\nu}{N}}
}
\nonumber \\&
=\sum_{k=-N/2}^{N/2-1}\frac{d_{k}^{(0)}}{\sqrt{N}}
\frac{
\sin\left(
\pi \left(\frac{N_{\mathrm{CP}}+n}{N}+\frac{1}{2}\right)(k-f+\nu)
\right)
}{e^{-\jmath \pi (k-f+\nu)\left(
\frac{n-N_{\mathrm{CP}}-1}{N}+\frac{1}{2}
\right)}
\sin\left(
\pi \frac{k-f+\nu}{N}
\right)
}\nonumber
\end{align}
where first, the formula for the sum of geometric progression, and then the Euler definition of sinus is used.
\paragraph{The calculation window spans only over the $0$-th NC-OFDM symbol, i.e., $n\in\{-N_{\mathrm{CP}},...,\frac{N}{2}\}$, giving}
\begin{align}
\label{eq:appendix5}
&b^{(0)}_{n}=\sum_{m=0}^{\frac{N}{2}-1}
\sum_{k=-N/2}^{N/2-1}
\frac{d_{k}^{(0)}}{\sqrt{N}}e^{\jmath 2\pi \frac{(n+m)k}{N}}e^{-\jmath 2 \pi (f-\nu) \frac{n+m}{N}}
 \\ &
=\sum_{k=-N/2}^{N/2-1}\frac{d_{k}^{(0)}}{\sqrt{N}}
\frac{
\sin\left(
\frac{\pi}{2}(k-f+\nu)
\right)
}{e^{-\jmath \pi (k-f+\nu)\left(
\frac{2n-1}{N}+\frac{1}{2}
\right)}
\sin\left(
\pi \frac{k-f+\nu}{N}
\right)
}
\nonumber
\end{align} 
by following the same steps as in (\ref{eq:appendix4}).
\paragraph{The calculation window spans partially over the end of the $0$-th NC-OFDM symbol, i.e., $n\in\{\frac{N}{2}+1,...,N-1\}$, giving}
\begin{align}
\label{eq:appendix6}
&b^{(0)}_{n}=\sum_{m=0}^{N-1-n}
\sum_{k=-N/2}^{N/2-1}
\frac{d_{k}^{(0)}}{\sqrt{N}}e^{\jmath 2\pi \frac{(n+m)k}{N}}e^{-\jmath 2 \pi (f-\nu) \frac{n+m}{N}}
 \\ &
=\sum_{k=-N/2}^{N/2-1}\frac{d_{k}^{(0)}}{\sqrt{N}}
\frac{
\sin\left(\pi
\frac{N-n}{N}(k-f+\nu)
\right)
}{e^{-\jmath \pi (k-f+\nu)\left(
\frac{n+N-1}{N}
\right)}
\sin\left(
\pi \frac{k-f+\nu}{N}
\right)
}
\nonumber
\end{align}
by following the same steps as in (\ref{eq:appendix4}).

Importantly, in all three cases, i.e., (\ref{eq:appendix4}), (\ref{eq:appendix5}), (\ref{eq:appendix6}), a given ($k$-th) NC-OFDM subcarrier waveform is \emph{sinc}-like with the same value of $\sin\left(\frac{k-f+\nu}{N}\right)$ in the denominator. The maximum of its absolute value for a given $k$ is obtained for $k-f+v=0$ in case b) and equals $|d_{k}^{(0)}\frac{\sqrt{N}}{2}|$. However, as the distance in the frequency between NC-OFDM active subcarriers (for which $|d_{k}^{(0)}|\neq 0 \forall k\in \mathbf{I}$, with the exception of the S\&C preamble, where subcarriers of even indices out of $\mathbf{I}$ are modulated with zeroes too) and the NBI frequency $f$ increase (assuming $|\nu|<1$), the envelope of $b_n^{(0)}$ decreases according to the factor of $1/|\sin\left(\frac{k-f+\nu}{N}\right)|$. For example, for $N\gg k-f+\nu$ the approximation $\sin(\pi\frac{k-f+\nu}{N})\approx \pi\frac{k-f+\nu}{N}$ holds, resulting in $|d_{k}^{(0)}\frac{\sqrt{N}}{(k-f+\nu)\pi}|$ being $\frac{\pi}{2}|k-f+\nu|$ times smaller than the peak of $|d_{k}^{(0)}\frac{\sqrt{N}}{2}|$. Moreover, the guard band of a few subcarriers causes a rapid decrease of $b_{n}^{(0)}$, which further results in negligible values of 
%$G_{\mathrm{yi}}(n)$, $G_{\mathrm{iy}}(n)$ 
$G_{\mathrm{cross}}(n)$
and $Q_{\mathrm{cross}}(n)$ in comparison to the standard OFDM system.

\begin{figure}[!t]
\centering
\includegraphics[width=3.4in]{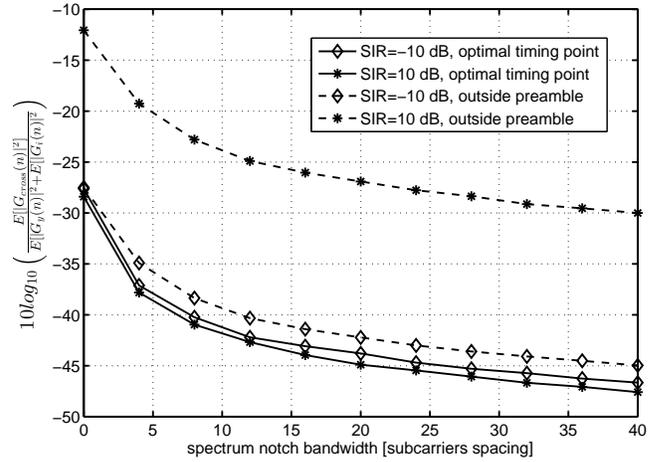}
\caption{Relative power of neglected term $G_{\mathrm{cross}}(n)$ vs spectrum notch bandwidth.}
\label{fig_neglected_term_power}
\end{figure}

The above is confirmed in Fig. \ref{fig_neglected_term_power} where the ratio of the mean power of component $G_{\mathrm{cross}}(n)$ to the sum of the mean powers of components $G_{\mathrm{y}}(n)$ and $G_{\mathrm{i}}(n)$ ($\frac{\mathbb{E}[|G_{\mathrm{cross}}(n)|^{2}]}{\mathbb{E}[|G_{\mathrm{y}}(n)|^{2}]+
\mathbb{E}[|G_{\mathrm{i}}(n)|^{2}]}$) is plotted. The results have been obtained by means of computer simulations for various spectrum notch bandwidths (around the perfect NBI frequency), various SIR values and two considered timing positions: the optimal timing point and a random point within one of NC-OFDM data symbols. It is visible that the mean power of the neglected term is much smaller than the sum of powers of the considered terms even for NBI overlapping in frequency with the utilized NC-OFDM subcarriers, i.e., there is no spectrum notch as in the standard OFDM case. However, as the spectrum notch around the NBI frequency increases, the relative power of $G_{\mathrm{cross}}(n)$ decreases further. It confirms that $b_{n}$ utilizing components, i.e., $G_{\mathrm{cross}}(n)$
and $Q_{\mathrm{cross}}(n)$, can be neglected in the NC-OFDM case. 

  % for no appendix heading


\ifCLASSOPTIONcaptionsoff
  \newpage
\fi



\bibliographystyle{IEEEtran}
\bibliography{pawla_bib}
%\begin{thebibliography}{1}
%\end{thebibliography}



%\vfill


\end{document}